\documentclass[%
 reprint,
 amsmath,amssymb,
 aps,
]{revtex4-2}

\usepackage{graphicx}
\usepackage{dcolumn}
\usepackage{bm}

\begin{document}


\title{The Hawking radiation in Massive Gravity: Path Integral and the Bogoliubov method}

\author{Ivan Arraut}
 \affiliation{FBL, University of Saint Joseph\\
  Estrada Marginal da Ilha Verde, 14-17, Macao, China.}
\author{Carlos Segovia}%
\affiliation{\quad Instituto de Matem\'aticas,\\
Universidad Nacional Autonoma de Mexico (IMUNAM), Oaxaca, M\'exico}%


\author{Wilson Rosado}
\affiliation{\quad Dpto. de Física, Universidad de Sucre\\ 
Cra. 28 No 5-267, Puerta Roja, Sincelejo-Sucre, Colombia}


\date{\today}

\begin{abstract}
We prove the consistency of the different approaches for deriving the black-hole radiation for the spherically symmetric case inside the theory of Massive Gravity. By comparing the results obtained by using the Bogoliubov transformations with those obtained by using the Path-Integral formulation, we find that in both cases the presence of the extra-degrees of freedom create the effect of extra-particles creation due to the distorsions on the definitions of time defined by the different observers at large scales. This however does not mean extra-particle creation at the horizon level. Instead, the apparent additional particles perceived at large scales, emerge from the way distant observers define their time coordinate, which is distorted due to the existence of extra-degrees of freedom.
\end{abstract}

\maketitle


\section{Introduction}

The theory of General relativity (GR) predicts the existence of black-holes. The classical theory suggests that no object can escape from a black-hole once it crosses the event horizon. In this way, although the thermodynamics of black-holes was developed a long time ago, it was believed by then that these objects cannot emit radiation \cite{a1,b1,b2,b3}. However, Hawking demonstrated in a seminal paper that quantum effects can make the black-holes to evaporate by emitting particles at a rate defined by the surface gravity \cite{a2,a3}. In its original derivation, Hawking used the method of Bogoliubov transformations in order to compare two different vacuums, one located at the future null infinity and the other one located at the past null infinity. The effect of particle creation then appeared as a consequence of the mix of positive and negative frequencies. The particle emission process was also proven by using the Path-Integral method where the periodicity of the poles of the propagators of a scalar field is equivalent to the effect of particle creation \cite{a4}. In this paper, we make the two derivations of the black-hole temperature as it is perceived by observers in Massive Gravity. In both methods, there appear modifications of the surface gravity due to the presence of the extra-degrees of freedom entering as a distortion of the notion of time in the theory.
Note that the modifications appear as a consequence of the way how the observers define the notion of time with respect to the preferred time-direction defined by the St\"uckelberg function $T_0(r, t)$, which contains the effect of the extra-degrees of freedom of the theory. This means that the fact that observers located at large scales in Massive Gravity define a different surface gravity with respect to observers in GR, does not mean that there is an extra-particle creation at the horizon level. Instead, what is happening in reality is that since the notion of particle depends on the way how we define the positive frequencies; and simultaneously, the definition of a positive frequency depends on the way how we define the time-coordinate, then the disagreement in the amount of particles between GR and Massive Gravity comes out from the fact that the extra-degrees of freedom affect and distort the time-coordinate in Massive Gravity. In other words, an observer in GR located at large scales from a black-hole will define the time in a different way in comparison with an observer located at the same scale but operating inside the theory of Massive Gravity. The definition of time naturally affects the definition of frequency. The paper is organized as follows: In Sec. (\ref{sec1}), we make a brief review about the most generic black-hole solution inside the scenario of Massive Gravity. In Sec. (\ref{Sec2}), we revise the Bogoliubov transformation method applied to Massive Gravity in order to calculate the amount of particles perceived by an observer located at a large distance with respect to the black-hole. In Sec. (\ref{Sec.4}), we develop the Path-Integral method for analyzing the same amount of particles perceived by an observer located at large scales with respect to the center of the black-hole. We then proceed to compare the results with those obtained via Bogoliubov transformations. Finally, in Sec. (\ref{Sec5C}), we conclude.

\section{The black-hole solution in Massive Gravity}   \label{sec1}

The black-hole solutions in Massive Gravity can be obtained after solving the field equations which come from the massive action
\begin{equation}\label{eq1}
S=\frac{1}{2\kappa}\int d^4 x\sqrt{-g}(R+m^2_gU(g,\phi))
\end{equation}
Here $U(g,\phi)$ is the potential term and it is defined as
\begin{equation}\label{eq2}
U(g,\phi)=U_2+\alpha_3U_3+\alpha_4U_4\,.
\end{equation}
Here $\alpha_3$ and $\alpha_4$ are the two free-parameters of the theory. The definitions for each order of the potential $U_n(g, \phi)$ can be found in \cite{a5,c1,c2}. The field equations are then
\begin{equation}\label{eq3}G_{\mu\nu} = -m_g^2X_{\mu\nu}\,,\end{equation}
with the energy-momentum tensor given by
\begin{equation}\label{eq4}X_{\mu\nu} = \frac{\delta U}{\delta g^{\mu\nu}}-\frac{1}{2}Ug_{\mu\nu}\,.\end{equation}
In eq. (\ref{eq3}), $m_g$ corresponds to the graviton mass. The spherically symmetric solutions for the previous field equations are obtained as
\begin{equation}\label{eq5}
ds^2=G_{tt}dt^2+G_{tt}S^2_0dr^2+G_{rt}(drdt+dtdr)+S^2_0r^2d\Omega^2_2\,, 
\end{equation}
where 
\begin{eqnarray}\label{eq6}
G_{tt}=-f(S_0r)(\partial_tT_0(r,t))^2,\nonumber\\
G_{rr}=-f(S_0r)(\partial_rT_0(r,t))^2
+\frac{1}{f(S_0r)},\nonumber\\
G_{tr}=-f(S_0r)\partial_tT_0(r,t)\partial_rT_0(r,t)\,.
\end{eqnarray}
Here
$f(S_0r)=1-\frac{2GM}{S_0r}-\frac{1}{3}\Lambda(S_0r)^2$, with $M$ being the mass of the black-hole, $\Lambda$ being a constant and $S_0$ being a parameter to be defined here. In a compact form, the spherically symmetric solutions, can be found to be 
\begin{equation}\label{eq7}
ds^2=-f(S_0r)dT_0(r,t)^2+\frac{S_0^2dr^2}{f(S_0r)}+S_0^2r^2d\Omega^2\,,
\end{equation}
working in unitary gauge \cite{a5}. In this metric, the St\"uckelberg function operates as a preferred
direction of time, different in general to the ordinary time-coordinate direction t. Then it
is necessary to define two different time-like Killing vectors; one in the direction $T_0(r, t)$
and another one pointing in the direction of the ordinary time-coordinate t. This mismatch
between the directions of the two Killing vectors, will generate a difference between the
amount of particles perceived by observers defined in Massive Gravity, and the amount of
particles perceived by observers satisfying the same conditions of motion in GR \cite{i1,i2}. In
general, it is known that
\begin{equation}\label{eq8}
T_0(r,t)=S_0t+A(r,t)\,.
\end{equation}
Here $T_0(r, t)$ behaves as a preferred time-direction \cite{i3,r1,r2}. $S_0$ is a scale factor depending on the
two free-parameters of the theory \cite{a5}. However, in this paper we sill focus on the case where the relation $\beta=\alpha^2$ is satisfied. This reduces the number of free-parameters to only one and then $S_0$ is defined as   

\begin{equation}
S_0=\frac{\alpha}{1+\alpha}.    
\end{equation}
Here the connection between $\alpha$ and $\beta$ with the two free-parameters $\alpha_3$ and $\alpha_4$ given in eq. (\ref{eq2}) is the following \cite{a5}

\begin{equation}
\alpha=1+3\alpha_3,\;\;\;\beta=3(\alpha_3+4\alpha_4).    
\end{equation}
Then in this paper we will analyze the particle creation process of the black-hole solution in Massive Gravity for the solution (\ref{eq7}) and for the case where we only have one free-parameter satisfying then the condition $\beta=\alpha^2$.

\section{The Bogoliubov transformation method in Massive Gravity: Hawking Radiation}   \label{Sec2}

In the Bogoliubov transformation method, we have to define a couple of vacuums. Both
vacuums will define a different St\"uckelberg function and as a consequence, a different value
for the function $A(r, t)$. Then both vacuums will be inequivalent in general. This inequivalence between the pair of vacuums under study can be perceived by the observers
in Massive Gravity as a particle creation effect. However, as it is the case when we analyzed
the Path-Integral formulation, the amount of particles emitted by the black-hole at the
event horizon does not change in this case with respect to the situation analyzed in GR.
However, the fact that the extra-degrees of freedom create the distortion effect, can make
the observers located at large scales to believe that there are extra-particles emitted by the
black-hole. In fact, this is just an illusion in the sense that there are no extra-particles
coming from the horizon. However, the effect is real in the sense that the distortion of
time is equivalent to a distortion of the notion of vacuum and then the observer's detectors
will really perceive an extra-component of radiation. The results obtained for the observers
defining the time coordinate in agreement with $T_0(r, t)$, will not differ with respect to the
results reported by observers in GR. Then we can use the standard and well-known Penrose
diagrams if we use the transformed St\"uckelberg functions $U(r, T_0(r, t)) = u + T_0(r, t)$ and
$V (r, T_0(r, t)) = v + T_0(r, t)$. Then the causal structure of the spacetime defined with respect
to $T_0(r, t)$ will be the same as in GR. Without loss of generality, we will take the spacetime
\begin{figure}[h!]
\centering
\includegraphics[scale=0.13]{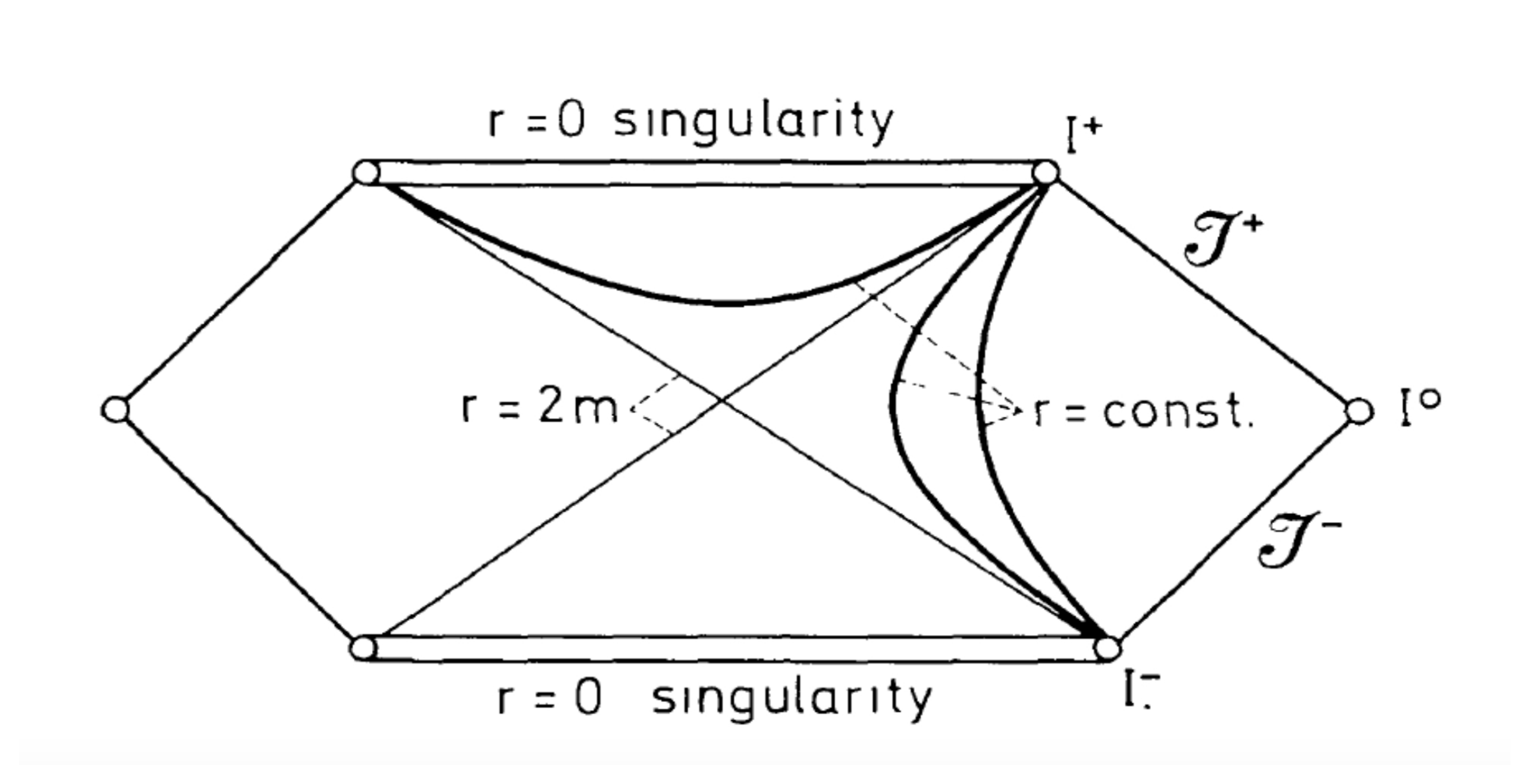} 
\caption{The Penrose diagram for the Schwarzschild geometry in GR as is shown in \cite{a2}. In Massive Gravity, the same diagram is valid if we express the black hole solution in terms of the St\"uckelberg functions. The past-null infinity (${J}^-$) of the diagram corresponds to the event where the black hole has not yet formed. The future null infinity (${J}^+$), on the other hand, corresponds to the case where the black hole is already formed. \label{fig1}}
\end{figure}
defined with respect to $T_0(r, t)$ as asymptotically flat. The asymptotically flat diagram can
be seen in the figure (\ref{fig1}). The deviations with respect to the usual notion of time due
to the presence of the extra-degrees of freedom have to be appreciable enough in order to
create distortions of time over the observers located at large scales and defining the time
arbitrarily. This distortion will affect the way how the particles are defined in the theory
of Massive Gravity and possibly the causal structure of spacetime. If we define on ${J}^+$ the
vector $n^a$, which is a future-directed null vector at $x$, pointing radially inward. Then the
vector $-\epsilon n^a$ joins the future event horizon with a surface of constant $U(r, T_0(r, t))$. 
We then define another null vector tangent to the horizon such that the normalization $n^a l_a = -1$ is
valid. Here we will demonstrate that the presence of the extra-degrees of freedom in massive
gravity, create a distortion which will modify the relation between the coordinate $u$ and the
affine parameter $-\epsilon$ \cite{a2,a3}. After doing the corresponding parallel transport of the vectors, we
obtain the relation
\begin{equation}\label{eq9}
\epsilon =Ce^{\kappa U(r,T_0)}\,,
\end{equation}
between the affine parameter and the St\"uckelberg function $U(r, T_0(r, t))$. In order to understand the
previous relation, we have to define the scalar field moving around the black-hole by expanding it in terms of positive and negative frequencies as
\begin{equation}\label{eq10}
\phi=\sum_i(f_i\hat{a}_i+\overline{f}_i\hat{a}_i^+)\,.
\end{equation}
Here the solutions for $f_i$ form a complete set of orthonormal functions over the past infinity
${J}^-$. Then they contain positive frequencies only with respect to the canonical affine parameter on ${J}^-$ \cite{a2,a3}. In the coming analysis we will need to define the orthonormal condition,
given by
\begin{equation}\label{eq11}
\frac{1}{2}\int_S(f_i\overline{f}_{j;a}-\overline{f}_jf_{i;a})d\Sigma^a=\delta_{ij}
\end{equation}
Here the integration is done over a suitable surface $S$. The upper bar over the functions
means complex conjugation operation. We remark that the functions $f_i$ have all the Cauchy data
defined over the past infinity. Then the operators $\hat{a}_i$ and $\hat{a}_i^+$ have the natural interpretation
of annihilation and creation operators for particles over the the past infinity (incoming
particles). We can define everywhere the field in the form given by its expansion with
respect to the functions $f_i$. However, it is also possible to expand the same field as follows
\begin{equation}\label{eq12}
\phi=\sum_i\left(p_i\hat{b}_i+\overline{p}_i\hat{b}_i^++q_i\hat{c}_i+\bar{q}_i\hat{c}_i^+\right)\,.
\end{equation}
Here the functions $p_i$ have zero Cauchy data at the future event horizon and they represent
outgoing components. They form an orthonormal family over the surface ${J}^+$ and they
only have positive frequencies with respect to the affine parameter along the null geodesic
generators on ${J}^+$. Then the operators $\hat{b}_i$ and $\hat{b}_i^+$ represent the annihilation and creation
operators for particles on the future infinity ${J}^+$. On the other hand, the functions $q_i$ have
zero Cauchy data at the future infinity ${J}^+$. They form a complete set of orthonormal
functions over the event horizon. However, it is not possible to define in this case a region
over which we can define positive frequencies and then the meaning of the operators $\hat{c}_i$ and $\hat{c}_i^+$
is not clear for this case, although not important at the moment of doing the calculations of
the black-hole radiation. What is important at this point is that the scalar field can either,
be expanded in terms of the functions $f_i$ or in terms of the functions $p_i$ and $q_i$. In order to
keep the canonical commutation relations, the previous functions, as well as the operators,
must be related to each other through the Bogoliubov transformations defined as follows
\cite{a2,a3}.

\begin{equation} \label{eq13}
\begin{split}
p_i & = \sum_j\left( \alpha_{ij}f_j  +\beta_{ij} \overline{f}_j  \right)\,, \\
 q_i & = \sum_j\left(   \gamma_{ij}f_j+\eta_{ij}\overline{f}_j   \right)\,,
\end{split}
\end{equation}
for the functions and 
\begin{equation} \label{eq14}
\begin{split}
\hat{b}_i& = \sum_j\left( \overline{\alpha}_{ij}\hat{a}_j  +\overline{\beta}_{ij} \hat{a}_j^+ \right)\,, \\
 \hat{c}_i & = \sum_j\left(   \overline{\gamma}_{ij}\hat{a}_j+\overline{\eta}_{ij}\hat{a}_j^+   \right)\,,
\end{split}
\end{equation}
for the annihilation operators. The creation operators can be obtained by applying the
adjoint operation over the previous operators. It is then clear that the fact that we have
an initial vacuum with no particles $\hat{a}_i|0\rangle = 0$, does not guarantee that other vacuums
defined in other locations of spacetime share the same definition of particle. This happens
when $\beta_{ij}\neq  0$, which is the coefficient mixing the positive and negative frequencies. Then
the observers located at the future infinity ${J}^+$ will perceive particle creation. In fact, the
expectation value of the number operator defined by using the operators $\hat{b}_i$ is 
\begin{equation}\label{eq15}
\langle0|\hat{b}_i^+\hat{b}_j|0\rangle=\sum_j|\beta_{ij}|^2
\end{equation}
 This just corresponds to the standard understanding of the concept of particle in asymptotically flat spacetimes \cite{e1}. In what follows we will divide the analysis in two parts, namely,
the case of GR which is equivalent to say that the observers in Massive Gravity take the
time coordinate as $T_0(r, t)$, and the other case is when the observers assume an arbitrary
direction for the time-coordinate. The way how observers define their local time depends on their conditions of motion.
\subsection{The case of GR: Observers defining the time in agreement with $T_0(r, t)$ in
Massive Gravity}

For this case, we can use safely the relation \eqref{eq9}. In such a case, it is easy to see that the
surfaces of constant phase $\omega U(r, T_0(r, t)$ are defined as
\begin{equation}\label{eq16}
\omega U=-\frac{\omega}{\kappa}(\log\epsilon-\log C)\,.
\end{equation}
On the past infinity ${J}^-$, the Killing vector $K^a$ is parallel to the vector $n^a$. Then we can
assume a relation $n^a = DK^a$, which just rescale the time coordinate and then the phase of
the solution in the past infinity is given by
\begin{equation}\label{eq17}
-\frac{\omega}{\kappa}\left(  \log(V_0-V)-\log D-\log C  \right)\,.
\end{equation}
Here again $V$ corresponds to the light-like St\"uckelberg function and it depends explicitly on $r$
and $T_0(r, t)$. The solutions corresponding to the Fourier components, expressed in spherically
symmetric form and in terms of advanced and retarded time (St\"uckelberg functions), are
defined as
\begin{equation}\label{eq18}
\begin{split}
f_{\omega',l,m} =&(2\pi)^{-1/2}r^{-1}(\omega')^{-1/2}F_{\omega'}(r)e^{i\omega'V}Y_{l,m}(\theta,\phi)\,,\\
p_{\omega,l,m}=&(2\pi)^{-1/2}r^{-1}(\omega)^{-1/2}P_\omega(r)e^{i\omega U}Y_{l,m}(\theta,\phi)\,.
\end{split}
\end{equation}
Here $Y_{l,m}(\theta, \phi)$ are the spherical harmonics normalized in the standard form. Due to the
Bogolibov transformations, we can express the functions $p_i$ as a linear combination of the
functions $f_i$ as has been explained previously. This is possible after doing the integration
over the frequency $\omega'$ as follows
\begin{equation}\label{eq19}
p_\omega=\int_0^\infty d\omega'\left(  \alpha_{\omega,\omega'} f_{\omega'} + \beta_{\omega,\omega'} \overline{f}_{\omega'}  \right)\,.
\end{equation}
Here we ignore the indices corresponding to the spherical harmonics, understanding that
formally they must appear. However, their presence will not contribute to the physics
developed in this section. If we replace the phase \eqref{eq17} inside the solution for the function $p_{\omega, l, m}$
defined in eq. \eqref{eq18}, then we get
\begin{equation}\label{eq20}
p_\omega^{(2)}\approx(2\pi)^{-1/2}r^{-1}(\omega)^{-1/2}P_\omega^-\left(\frac{V_0-V}{CD}\right)^{-i\frac{\omega}{\kappa_{eff}}}\,.
\end{equation}
If we make a Fourier transformation with respect to $V$ , it is trivial to demonstrate that the
Bogoliubov coefficients $\alpha_{\omega,\omega'}^{(2)}$ and $\beta_{\omega,\omega'}^{(2)}$ are defined for large values of $\omega'$ as
\begin{equation}\label{eq21} 
\begin{split}
 \beta_{\omega,\omega'}^{(2)}\approx&   - i\alpha_{\omega,(-\omega')}^{(2)} \,,\\
  \alpha_{\omega,\omega'}^{(2)}\approx &(2\pi)^{-1}P_\omega^-(CD)^{i\frac{\omega}{\kappa_{eff}}} e^{i(\omega-\omega')v_0} \left(\frac{\omega'}{\omega}\right)^{1/2}\times\\
 &\Gamma\left( 1-\frac{i\omega}{\kappa}\right)(-i\omega')^{-1+i\frac{\omega}{\kappa_{eff}}}
\end{split}
\end{equation}
Here $\alpha^{(2)}$ can be obtained from $\beta^{(2)}$ if we make an analytical continuation of $\beta^{(2)}$ around
the logarithmic singularity defined by the term 
$(-i\omega')^{-1+i\frac{\omega}{\kappa_{eff}}}$
in the previous results. In fact,
if we round the logarithmic singularity in the neighborhood of $\omega'\rightarrow 0$ by taking $\omega'\rightarrow e^{i\pi}\omega'$,
the we obtain
\begin{equation}\label{eq22}
|\alpha^{(2)}_{\omega,\omega'}|=e^{\frac{\pi\omega}{\kappa_{eff}}}|\beta^{(2)}_{\omega,\omega'}|\,.
\end{equation}
This result corresponds to the standard one derived by Hawking \cite{a2,a3} and it will define the
amount of particles which an observer defining the time in agreement with $T_0(r, t)$, will
perceive. In this part of the paper, we have defined the surface gravity as $\kappa_{eff}$. This quantity will
be defined as the effective surface gravity perceived by the observers defining the time in
agreement with $T_0(r, t)$. Observers defining the time in a different way, will perceive a different $\kappa$ which can be connected functionally with $\kappa_{eff}$. 

\subsection{The case of observers defining the time arbitrary}
In this subsection we will derive the surface gravity $\kappa$ for observers defining the time arbitrarily by calculating first the effective surface gravity $\kappa_{eff}$ as a function of $\kappa$. Once again we remark that $\kappa_{eff}$ corresponds to the standard result perceived by the observers defining the time $T_0(r,t)$. We can repeat the previous reasoning
for the case of observers defining the time arbitrarily, then some changes will appear due to
the presence of the function $A(r, t)$ inside the advanced and retarded light-like St\"uckelberg
functions $U(r, T_0(r, t))$ and $V (r, T_0(r, t))$. Here $T_0(r, t)$ is the standard St\"uckelberg function.
By taking into account that $T_0(r, t) = S_0t + A(r, t)$ and rescaling the time-coordinate as
$t\rightarrow S_0t$ \cite{a5}; then the surfaces of constant phase defined previously become
\begin{equation}\label{eq23}
\omega u = -\frac{\omega}{\kappa}(\log \epsilon-\log C)-\omega A(r,u)\,.
\end{equation}
This redefinition of the surfaces of constant phase, depends on $A(r, u)$ which is the distortion
of time created by the extra-degrees of freedom at large scales. This function $A(r, u)$, depends explicitly on the variable u. This point will be important at the moment of calculating the Black-hole radiation perceived
by the observers defining the time arbitrarily. Note that it is expected the function $A(r, u)$
to vanish in the neighborhood of the future event horizon if GR is recovered for strong
gravitational fields as the standard theory of Massive Gravity suggests. We can express the contribution of $A(r, u)$ in a different way, such that
the result looks like
\begin{equation}\label{eq24}
\omega u=-\frac{\omega}{\kappa}(\log \epsilon-\log C-\log (e^{\kappa A(r,u)}))\,.
\end{equation}
Here again on the past infinity, ${J}^-$, the Killing vector $K^a$
is parallel to the vector $n^a$. Then
we can assume the same relation $n^a = DK^a$ and then we get
\begin{equation}\label{eq25}
-\frac{\omega}{\kappa}(\log (V_0-V))-\log D-\log C-\log (e^{\kappa A(r,u)})\,,
\end{equation}
in close analogy to what happens in \cite{a2,a3}. In eq. (\ref{eq25}), $V$ is defined as $V = v + A(r, v)$. However, near the past event horizon, it is also expected the extra-degrees of freedom to
become negligible and then $V\approx  v$. This approximation is not valid at scales far away from
the past event horizon and then we will keep in mind the fact that we have to include the
function $A(r, v)$ in the calculations. A detail to remark is that here we are taking $A(r, u)$ as
the function related to $U(r, T_0(r, t)) = u + A(r, u)$. This is the case because here we define
the retarded time as given by $u$. On the other hand, we also define $V (r, T_0(r, t)) = v+A(r, v)$
by defining in this case the advanced time as $v$. The relation between $u$ and $v$ is defined by
the result \eqref{eq17}. Along the past infinity, the solution will be
\begin{eqnarray}\label{eq26}
p_\omega^{(2)}\sim (2\pi)^{-1/2} \omega^{-1/2} r^{-1}P_\omega^-\times\nonumber\\
exp\left( -\frac{i\omega}{\kappa}\log \left(  \frac{v_0-v- A(r,v)}{CDe^{\kappa A(r,u)}}   \right)  \right)
\end{eqnarray}
The difference between the standard case and the Massive Gravity one is the presence of the
term $A(r, u)$, which will affect the integration over the variable $v$ at the moment of doing the
Fourier transformation in order to find the Bogoliubov coefficients. Later we will see that the
term $A(r, v)$ does not affect the integration over $v$ after doing the appropriate substitution
of variables. The case of $A(r, u)$ is different due to the non-trivial relation between $u$ and
$v$. However, the fact that we still have a black-hole emitting particles in agreement with
the Fermi-Dirac statistic remain. The fraction of particles entering the black hole is given by \cite{a2,a3}
\begin{equation}\label{eq27}
\Gamma_{jn}=\int_0^\infty\left( |\alpha_{\omega,\omega'}^{(2)}|^2-|\beta_{\omega,\omega'}^{(2)}|^2   \right)\,.
\end{equation}
In order to find the Bogoliubov coefficient $\alpha^{(2)}$, then we have to multiply the result \eqref{eq27} by
$\overline{f}_{j;a}$ here defined by
\begin{eqnarray}\label{eq28}
\overline{f}_{\omega';v}=-i(1+\partial_vA(r,v))(2\pi)^{-1/2}r^{-1}(\omega')^{1/2}\times\nonumber\\
F_{\omega'}(r)e^{-i\omega'(v+A(r,v))}\,.
\end{eqnarray}
This result is obtained from eq. \eqref{eq18} if we take the complex conjugate for the function $f_{\omega'}$
and we take the derivative with respect to $v$. Here we have used the definition $V = v+A(r, v)$ and we
have also ignored the spherical harmonics contribution which we assume to be normalized in the
standard way. By multiplying the result \eqref{eq26} with the previous result and then integrating
over the variable $v$ in order to make the Fourier transformation, we obtain
\begin{equation}\label{eq29}
\begin{split}
&\alpha_{\omega,\omega'}^{(2)}\approx -(2\pi)^{-1} P_\omega^-(CD)^{i\frac{\omega}{\kappa}} \left( \frac{\omega'}{\omega} \right)^{-1/2} \times\\
&\int (v_0-v-A(r,v))^{-i\frac{\omega}{\kappa}}e^{i\omega A(r,u)} (1 + \partial_v A(r,v)) e^{i\omega'(v+A(r,v))}dv\,.
\end{split}
\end{equation}
Here $V_0\approx v_0$ since $v_0$ has a correspondence with the past event horizon by assuming that the whole Penrose diagram corresponds to the vacuum Schwarzschild solution. Here
the integration is done for all the values of $v$. In this case, in general, we do not get
the same Gamma function obtained previously inside the standard calculations. Instead, the result in eq. (\ref{eq29}) depends on the functional
behavior of $A(r, u)$. It is trivial to observe that the result will not depend on $A(r, v)$. If we make the
replacement $z = v_0 -v -A(r, v)$, in the previous integral, then we get
\begin{equation}\label{eq30}
\begin{split}
&\alpha_{\omega,\omega'}^{(2)}\approx -(2\pi)^{-1} P_\omega^-(CD)^{i\frac{\omega}{\kappa}} \left( \frac{\omega'}{\omega} \right)^{-1/2} \times\\
&\int (z)^{-i\frac{\omega}{\kappa}}e^{i\omega A(r,u(z))} e^{-i\omega'(v_0-z)}dz\,.
\end{split}
\end{equation}
Here we make one additional replacement by taking $x = -i\omega'z$, getting then the result
\begin{equation}\label{eq31}
\begin{split}
&\alpha_{\omega,\omega'}^{(2)}\approx -(2\pi)^{-1} P_\omega^-(CD)^{i\frac{\omega}{\kappa}}  e^{-i\omega'v_0} \left( \frac{\omega'}{\omega} \right)^{-1/2} \times\\
&  (-i\omega')^{-1+i\frac{\omega}{\kappa}} \int (x)^{-i\frac{\omega}{\kappa}}e^{i\omega A(r,u(x))} e^{-x}dx\,.
\end{split}
\end{equation}
For the case when $A(r, u) = 0$ we recover the result \eqref{eq21}. Here however, we will take the
function $A(r, u)$ as a polynomial expansion on the variable $u$ as follows
\begin{equation}\label{eq32}
A(r,u)\approx \sum_{n=0}^\infty a_n u^n\,.
\end{equation}
In addition we have to take into account the relation between $u$ and $v$ already obtained in
eq. \eqref{eq17}. Then here we use
\begin{equation}\label{eq33}
A(r,u)\approx \sum_{n=0}^\infty a_n \left( -\frac{1}{\kappa}    \log  \left(   \frac{v_0-v-A(r,v)}{DC}  \right) \right)^n
\end{equation}
Then the contribution of $A(r, u)$ to the integral in eq. \eqref{eq31} is
\begin{equation}\label{eq34}
e^{i\omega A(r,u(x))}=\left(   \frac{v_0-v-A(r,v)}{CD}   \right)^{i\omega\Sigma_n \frac{na_n}{\kappa^n}}\,.
\end{equation}
If we make the same changes of variables as before, this previous expression is given by
\begin{equation}\label{eq35}
e^{i\omega A(r,u(x))}=(-i\omega')^{i\omega\Sigma_n\frac{na_n}{\kappa^n}}(DC)^{i\omega\Sigma_n\frac{na_n}{\kappa^n}}(x)^{-i\omega\Sigma_n\frac{na_n}{\kappa^n}}
\end{equation}
In the previous series, the coefficient $a_1$, corresponding to $n=1$ can be defined in a convenient
way, such that it is possible to write the result (\eqref{eq31}) as
\begin{equation}\label{eq36}
\begin{split}
&\alpha_{\omega,\omega'}^{(2)}\approx -(2\pi)^{-1} P_\omega^-(CD)^{i\omega\sum_n\frac{na_n}{\kappa^n}}  e^{-i\omega'v_0} \left( \frac{\omega'}{\omega} \right)^{-1/2} \times\\
&  (-i\omega')^{-1+i\omega\sum_n\frac{na_n}{\kappa^n}} \int (x)^{-i\omega\sum_n\frac{na_n}{\kappa^n}} e^{-x}dx\,.
\end{split}
\end{equation}
The result is simply given by
\begin{equation}\label{eq37} 
\begin{split}
 \beta_{\omega,\omega'}^{(2)}\approx&   - i\alpha_{\omega,(-\omega')}^{(2)} \,,\\
  \alpha_{\omega,\omega'}^{(2)}\approx &(2\pi)^{-1}P_\omega^-(CD)^{i\omega\sum_n\frac{na_n}{\kappa^n}} e^{i(\omega-\omega')v_0} \left(\frac{\omega'}{\omega}\right)^{1/2}\times\\
 &\Gamma\left( 1-i\omega\sum_n\frac{na_n}{\kappa^n}\right)(-i\omega')^{-1+i\omega\sum_n\frac{na_n}{\kappa^n}}
\end{split}
\end{equation}
Here the relation between $\alpha^{(2)}$ and $\beta^{(2)}$ is not modified. In fact, if we make the
analytical continuation around the logarithmic singularity defined this time by the term
$(-i\omega')^{-1+i\omega\sum_n\frac{na_n}{\kappa^n}}$, in order to get $\alpha^{(2)}$ and $\beta^{(2)}$ by using $\omega'\rightarrow e^{i\pi}\omega'$, then we obtain the
relation
\begin{equation}\label{eq38}
|\alpha_{\omega,\omega'}^{(2)}|=e^{\pi\omega\sum_n\frac{na_n}{\kappa^n}}|\beta_{\omega,\omega'}^{(2)}|\,.
\end{equation}
This has a correspondence to the result \eqref{eq22}. Then the total number of particles created in
a given mode is defined by the relation
\begin{equation}\label{eq39}
|\beta_{\omega,\omega'}^{(2)}|\approx \Gamma_{\omega'}\left(  e^{2\pi\omega\sum_n\frac{na_n}{\kappa^n}}-1 \right)^{-1}\,,
\end{equation}
still consistent with the statistics followed by the black body radiation. From this previous
result however, we can define an effective surface gravity given by
\begin{equation}\label{eq40}
\kappa_{eff}=\left( \sum_n\frac{na_n}{\kappa^n} \right)^{-1}=\frac{1}{4GM}\,.
\end{equation}
Here $\kappa$ is the surface gravity perceived by an observer defining the time in an arbitrary way.
On the other hand, $\kappa_{eff}$ is the surface gravity perceived by an observer defining the time in
agreement with $T_0(r, t)$. Here we have used some specific functional dependence
for $A(r, u)$ and the result can change depending how this function behave. The method
developed is however general and the functional behavior selected is a polynomial expansion
which is what should be expected. The radial dependence of this function is irrelevant since
this variable will not enter in the integration over $v$. If we want to find the surface gravity
perceived by the observers defining the time arbitrarily, we have to solve eq. \eqref{eq40} for $\kappa$. It is evident that the result will be different to the standard one. Before moving forward with the Path-Integral methods, we must remark that the Bogoliubov method studies the gravitational collapse of the black-hole. This is the case because the past null infinity ${J}^-$ represents the vacuum state when there is no black hole or the standard vacuum state without particles. However, the future null-infinity ${J}^+$ represents the case where the black-hole is already formed and although locally an observer can still define a vacuum state, when we compare the vacuum state at ${J}^-$, with the vacuum state at ${J}^+$, then we find that there are certain amount of particles emerging at ${J}^+$ for the corresponding observers. It is for this reason that the Penrose diagrams are representations of the gravitational collapse.

\section{The Path Integral Formulation of the Black-Hole radiation in Massive Gravity}   \label{Sec.4}

The Path-Integral formulation for evaluating the black-hole temperature in Massive Gravity, was developed in \cite{i1,i2}. The result suggested that the periodicity of the poles of the
propagator for the scalar field is affected by the presence of the extra-degrees of freedom.
This is consistent with the fact that $T_0(r, t)$ behaves as a preferred direction of time and in
general, the analytical extension of $T_0(r, t)$, will differ from the analytical extension defined
for the ordinary time-coordinate t. The condition of regularity for the Cauchy data means
that the following result must be satisfied \cite{i1,i2}
\begin{equation}\label{eq41}
-4\pi G M<\psi(r,t) <0\,.
\end{equation}
Here $\psi (r, t) = \mu+\overline{A}(r, t)$, where we have defined the time coordinate as $t = \gamma+i\mu$, separating
it in real and imaginary part. In addition $A(r, t) = Re(A(r, t)) + i\bar{A}(r, t)$, having this
function real and imaginary part as well. We can notice that the imaginary part of the time-coordinate
defines the periodicity of the propagator. The periodicity will be affected for the observers
defining the time arbitrarily, if the function $A(r, t)$ has an imaginary component, namely, if
$\bar{A}(r, t) = 0$. The result \eqref{eq41} can then be expressed more explicitly as
\begin{equation}\label{eq42}
-4\pi GM<\mu+\bar{A}(r,t)<0\,.
\end{equation}
Then the amount of particles perceived by an observer in Massive Gravity, will depend on
how he/she defines the local time. If an observer defines the time in agreement
with $T_0(r, t)$, then the amount of particles perceived will be the same as in the GR case.
The temperature perceived by the observers defining an arbitrary direction of time $t$, will
depend on the explicit solution for $\mu$ coming from the condition of periodicity of the poles of
the propagator
\begin{equation}\label{eq43}
8\pi GM =\mu+\bar{A}(r,t)\,.
\end{equation}
Then the problem is reduced in finding the solution for $\mu$ from this previous expression.
Here $\bar{A}(r, t)$ can have any dependence. However, without loss of generality, we can take the
function $\bar{A}(r, t)$ to be a polynomial expansion of $\mu$, $\bar{A}(r, t) = \sum_{n=0}^{\infty}b_n\mu^n$. Here again the
linear term in the expansion can absorb the linear term µ and then we can express the result
as
\begin{equation}\label{eq44}
8\pi GM=\sum_{n=0}^\infty b_n\mu^n=\mu_{eff}\,.
\end{equation}
Here $\mu_{eff}$ is the complex component of time defined by the observers defining the time in
agreement with $T_0(r, t)$. On the other hand, $\mu$ is related to the observers defining the time
arbitrarily. Then we can define the black-hole temperature as
\begin{equation}\label{eq45}
T_{eff}=\frac{1}{\mu_{eff}}=\frac{1}{8\pi GM}=\frac{1}{\sum_nb_n\mu^n}\,.
\end{equation}
In this way, we can see that the observers defining the time arbitrarily will in general perceive a
different temperature with respect to the standard one calculated in GR. Order by order,
there is a direct correspondence between this previous result and the one obtained in eq.
\eqref{eq40}. This can be seen after taking into account the well-known relation between surface gravity and temperature here expressed as
\begin{equation}\label{eq46}
T=\frac{1}{2\pi}\kappa=\frac{1}{\mu}\,.
\end{equation}
This result is the temperature perceived by observers defining the time arbitrarily. $\mu$ can be
found by solving the polynomial equation \eqref{eq44}. The solution is in general non-trivial. Eq. (\ref{eq45}) has a direct relation with the expansion done in eq. (\ref{eq40}), where the Bogoliubov methods were used.

\subsection{Further analysis}
In order to see that the method used here is general, we will analyze the
functional dependence of $\bar{A}(r, \gamma + i\mu)$. In \cite{i1, i2} it was assumed that it was always possible to
find $\bar{A}(r, t)$ and then the surface gravity was defined for the simplest case. Here however
we go deeper in the analysis in order to explain that our previous result is general. For
simplicity, in order to illustrate the consistency, we will assume that $\mu << t$, namely, that
the imaginary component of the analytically extended time-coordinate is much smaller than
the real component. In such a case we can define the expansion
\begin{equation}\label{eq47}
\begin{split}
A(r,\gamma+i\mu)\approx &\sum_{n=even}\left(  n! \partial_{i\mu}^{(n)} A(r,t)_{\mu=0}\right) \cos \mu\\
&+ i\sum_{n=odd} \left( n! \partial_{i\mu}^{(n)} A(r,t)_{\mu=0}  \right) \sin\mu
\end{split}
\end{equation}
Then in this case we define the analytically extended St\"uckelberg function as
\begin{equation}\label{eq48}
\bar{A}(r,t)\approx\sum_{n=odd}\left( n!\partial_{i\mu}^{(n)} A(r,t)_{\mu=0}\right)\sin\mu\,.
\end{equation}
From this definition it is evident that the analytical extension of $T_0(r, t)$ will be different
to the analytical extension of $t$. This is the mismatch which makes the concept of particle
ambiguous in Massive Gravity for observers located at large scales. This distortion effect
is absent in GR and then it is a consequence of the extra-degrees of freedom. Whenever
the time is distorted, the notion of particle is modified with respect to the same observers
defined in GR. This will generate the effect of particle creation for observers located at
large scales in Massive Gravity. However, this does not mean that there are extra-particles
coming from the event horizon because the effects come from the ambiguity on the vacuum definitions generated by the extra-degrees of freedom of the theory. This is an interesting effect and it helps us to understand
that not only the curvature effects make it possible to perceive Hawking radiation, but also any
other contribution able to create distortions in the concept of time, will generate Hawking
radiation. This effect is general and it might appear whenever there are degrees of freedom or any other physical effect able to affect the way how we define the time at different scales. Going further into the previous calculations,
we can then express the analyticity condition for the propagator as
\begin{equation}\label{eq49}
-4\pi GM < \mu +\sum_{n=odd}\left( n!\partial_{i\mu}^{(n)} A(r,t)_{\mu=0}\right)\sin\mu <0
\end{equation}
The associated periodicity condition will be
\begin{equation}\label{eq50}
8\pi GM = \mu+ \sum_{n=odd}\left( n!\partial_{i\mu}^{(n)} A(r,t)_{\mu=0}\right)\sin\mu\,.
\end{equation}
If we define the temperature for an observer taking the time coordinate as $t$, then such
observer will define its imaginary time coordinate as $\mu$. In such a case, then the surface
gravity (temperature) perceived by the observer will be given by eq. \eqref{eq46}, with $\mu$ defined as
the solution of the equation
\begin{equation}\label{eq51}
y=8\pi GM-  \sum_{n=odd}\left( n!\partial_{i\mu}^{(n)} A(r,t)_{\mu=0}\right)\sin\mu =\mu\,.
\end{equation}
The solution for this equation is the intersection of the straight line $y = \mu$ and the function
$y=8\pi GM-  \sum_{n=odd}\left( n!\partial_{i\mu}^{(n)} A(r,t)_{\mu=0}\right)\sin\mu$. It is then evident that the temperature perceived by an observer defining the time in agreement with $t$, will differ from the one defined
by observers taking the time $T_0(r, t)$.

\section{Conclusion}   \label{Sec5C}

In this paper we have calculated, by using two different methods, the black-hole temperature for a spherically symmetric black-hole in the non-linear formulation of massive
gravity. For illustration, we have focused on the solution satisfying the relation $\beta=\alpha^2$. Then we reduce the theory to one free-parameter. Interestingly, we have found that both methods, namely, the Path-Integral method and the
Bogolibov transformation one, provide results consistent with each other. The results
suggest that the observers defining the time-coordinate in agreement with the St\"uckelberg
function $T_0(r, t)$, will perceive a temperature equivalent to the one observed in GR. On
the other hand, the observers defining the time arbitrarily $t$, will perceive a different
value of temperature with respect to the one perceived in GR with equivalent
conditions of motion. This happens at large scales where the effects of the distortion of
the time-coordinate generated by the extra-degrees of freedom become appreciable. The
distortion of the time-coordinate affects the way how the observers defining the coordinate $t$
perceive the periodicity of the poles of the propagator if they use the Path-Integral method
in their calculations. Equivalently, from the perspective of the Bogoliubov method, the
distortion of the time-coordinate affects the relation between the advanced and the retarded
time-coordinates when we relate the vacuum over the future infinity (${J}^+$) with the vacuum defined over the past null infinity (${J}^-$) in the analytically
extended Penrose diagrams, which are descriptions of the process of gravitational collapse. This is the case because what the Penrose diagrams portrait is: 1). The vacuum state before the formation of the black-hole (${J}^-$). 2). The vacuum state after the formation of the same black-hole (${J}^+$). The mentioned distortion of the time-coordinate generated by the extra-degrees of freedom in Massive Gravity, then affects the final result obtained for the coefficient $\beta_{\omega, \omega'}$, which is obtained after comparing the mentioned vacuum states. The same coefficient mixes the positive and negative frequencies. The effect described here is not standard and it does
not correspond to the emission of extra-particles from the event horizon. At the event
horizon scale, it is still expected that the black-holes in Massive Gravity emit the same amount
of particles as in GR. The effect discussed in this paper rather appears as a consequence of the ambiguity on the definitions of vacuum for the observers located at large scales with respect to the black-hole. The definitions of vacuum, naturally, depend on how the time-coordinate is defined for each observer.






\begin{thebibliography}{0}

\bibitem{a1} J. M. Bardeen, B. Carter, S. W. Hawking, 
The four laws of black hole mechanics.
Comm. Math. Phys. 31 (1973), 161-170. 

\bibitem{b1} J. D. Bekenstein, Black Holes and Entropy, Physical Review D, vol. 7, no. 8, (1973) 2333-2346.

\bibitem{b2}  J. D. Bekenstein, Generalized second law of thermodynamics in black-hole physics, Phys. Rev. D9, no. 12, (1974) 3292-3300.

\bibitem{b3} J. D. Bekenstein, Universal Bound on the Entropy to Energy Ratio for Bounded Systems. Phys. Rev. D23:  (1981)  287-298.

\bibitem{a2} S. W. Hawking, 
Particle creation by black holes.
Comm. Math. Phys. 43 (1975), no. 3, 199-220; 
\bibitem{a3}
S. W. Hawking, 
Erratum: "Particle creation by black holes''(Comm. Math. Phys. 43 (1975), no. 3, 199-220).
Comm. Math. Phys. 46 (1976), no. 2, 206. 

\bibitem{a4}  J. B. Hartle and S.W. Hawking, 
Path-Integral derivation of black-hole radiance.
Phys. Rev. D 13 (1976), no. 8, 2188-2203.

\bibitem{a5} H. Kodama and I. Arraut, Stability of the Schwarzschild-de Sitter black hole in the dRGT Massive Gravity theory, Progress of Theoretical and Experimental Physics, Volume 2014, Issue 2, 1 February 2014, 023E02.

\bibitem{c1} C. de Rham, G. Gabadadze and A. J. Tolley, Resummation of Massive Gravity, Phys. Rev. Lett. 106, 231101 (2011).

\bibitem{c2} C. de Rham and G. Gabadadze, Generalization of the Fierz-Pauli Action, Phys. Rev. D 82, 044020 (2010).

\bibitem{i1} I. Arraut, On the apparent loss of predictability inside the de-Rham-Gabadadze-Tolley non-linear formulation of Massive Gravity: The Hawking radiation effect, EPL 109 (2015) no. 1, 10002.

\bibitem{i2} I. Arraut, Path-Integral derivation of black-hole radiance inside the de-Rham-Gabadadze-Tolley formulation of Massive Gravity, I. Eur. Phys. J. C (2017) 77: 501.

\bibitem{i3} I. Arraut, The Black Hole Radiation in Massive Gravity, Universe 2018, 4(2), 27.

\bibitem{r1} V. Rubakov, P. G. Tinyakov, Infrared-modified gravities and massive gravitons, PHYS-USP, 2008, 51 (8), 759?792.

\bibitem{r2} V. Rubakov, Lorentz-violating graviton masses: getting around ghosts, low strong coupling scale and VDVZ discontinuity, arXiv:hep-th/0407104.
 
\bibitem{e1} N. D. Birrell, Nicholas David Birrell, P. C. W. Davies, Quantum Fields in Curved Space, Cambridge University Press, 1984.

\end{thebibliography}

\end{document}